\documentclass[preprint,amsmath,amssymb,amsfonts,aps,superscriptaddress,nofootinbib]{revtex4}
\usepackage{ifsym}
\usepackage{MnSymbol}
\usepackage[usenames,dvipsnames]{xcolor}
\usepackage[bookmarksnumbered]{hyperref}
\def\be{\begin{equation}}
\def\ee{\end{equation}}
\def\ba{\begin{eqnarray}}
\def\ea{\end{eqnarray}}
\def\nl{\nonumber\\}
\def\nn{\nonumber}

\def\a{\alpha}
\def\b{\beta}
\def\d{\delta}

\def\s{{\sigma}}

\def\l{\lambda}

  \def\r{\rho}

\def\cA{{\cal A}}

\def\[{\left[}
\def\]{\right]}
\def\({\left(}
\def\){\right)}

\def\<{\langle}
\def\>{\rangle}

\def\Tr{\text{Tr\,}}
\def\sun{{SU(N)}}  \def\son{SO(N)} \def\sp2n{Sp(2N)}

\def \bA{{\bar A}}

\def\2F1{\,_2{\rm F}_1}

\newcommand{\tr}{\text{Tr}}

\begin{document}

\title{Group-theoretic relations for amplitudes in gauge theories with orthogonal and symplectic groups}


\author{Jia-Hui Huang}
\email{huangjh@m.scnu.edu.cn}
\affiliation{Guangdong Provincial Key Laboratory of Quantum Engineering and Quantum Materials,
School of Physics and Telecommunication Engineering,
South China Normal University, Guangzhou 510006,China}


\date{\today}

\begin{abstract}
It is important to find  nontrivial constraint relations for color-ordered amplitudes in gauge theories. In the past several years, a pure group-theoretic iterative method has been proposed to derive linear constraints on color-ordered amplitudes in $\sun$ gauge theories. In this paper, we use the same method to derive linear constraints on four-point gluon amplitudes in $\son$ and $\sp2n$ gauge theories. These constraints are derived up to four-loop order. It is found that there are $n=1,6,10,13,16$ constraint relations at $L=0,1,2,3,4$ loop orders in both $\son$ and $\sp2n$ cases. Correspondingly, the numbers of independent four-point color-ordered amplitudes are $2,3, 5, 8, 11$ at $L=0,1,2,3,4$ loop orders in both theories.
\end{abstract}

\maketitle

\section{Introduction}\label{sec:intro}
Scattering amplitudes are important gauge invariant observables in perturbative quantum gauge theories. Many efforts have been made to simplify the computation of the color-ordered amplitudes in recent years. It is also found that color-ordered amplitudes are not all independent. Tree level color-ordered amplitudes in $\sun$ gauge theories satisfy many constraint relations, such as $U(1)$ decoupling relations \cite{bg1987,mpx1988,bernk1991,FHJ2011,hhj2011}, Kleiss-Kuijf(KK) relations \cite{kk1989,ddm2000} and Bern-Carrasco-Johansson(BCJ) relations \cite{bcj2008,bdv2009,Stieberger:2009hq,cdf2011,Mafra:2011kj}. The KK relations reduce the number of independent $n$-point color-ordered amplitudes with a given helicity configuration to $(n-2)!$. And the BCJ relations further reduce this number to $(n-3)!$.

Just as tree level color-ordered amplitudes, loop amplitudes in $\sun$ gauge theories are also not all independent to each other.
The loop-level $U(1)$ decoupling relations are well-known relations between loop amplitudes and exist at arbitrary loops \cite{bernk1991}.  These relations can be derived from the fact that $U(1)$ photon decouples from $\sun$ gluon scattering processes. The extension of KK and BCJ relations to one- and two-loop has also been explored \cite{ddm2000,bfd2002,cj2011,fjh2012,bi2012, dl2013}. In the past several years, some nontrivial all loop relations have been derived for four-, five- and six-point color-ordered loop amplitudes in $\sun$ gauge theories based on a  group-theoretic method \cite{Naculich:2011fw,nacu2012,enacu20121,enacu20122}.

So far color-ordered amplitudes in  $\sun$ gauge theories are studied extensively. One important reason is that $\sun$ gauge symmetry plays important role in modern standard model of particle physics. But it is known that the standard model suffers from several difficulties, such as neutrino oscillations and dark matter problems. So beyond the standard model, many candidate theories have been explored, like grand unified theories (GUT) and string theories. In these candidate theories, orthogonal gauge groups often play important roles, such as $SO(10)$ in GUT \cite{am1983,Babu:1992ia,Lazarides:1980nt,ckn1982,mkfn2002} and $SO(32)$ in string theories \cite{Green:1984ed,witten1984,Clavelli:1985me,nrvw2006,akot2015}. Thus, it is interesting to study scattering amplitudes in gauge theories with $\son$ or other gauge groups.

In this paper, we consider constraint relations for four-point color-ordered amplitudes in two kinds of gauge theories,
 namely, $\son$ gauge theories  and  $\sp2n$ gauge theories. All external and internal particles are in adjoint representations of the gauge groups. By employing the group-theoretic approach \cite{nacu2012,enacu20121,enacu20122}, we derive constraint relations satisfied by four-point color-ordered amplitudes in both gauge theories up to four-loop order.
It is found that, at $L=0,1,2,3,4$ loop orders, there are respectively $n=1,6,10,13,16$ group-theoretic constraint relations for four-point color-ordered amplitudes  in both kinds of theories. The number of independent color-ordered amplitudes at each loop order for both theories is listed in Table 1. These results are consistent with those from a direct counting of color basis elements and trace basis elements up to four loop \cite{Bern:2010tq,nacu2012}.
\begin{table}
\caption{Loop orders vs. Numbers of independent amplitudes}
\centering
\begin{tabular}{|p{1.5cm}|p{0.8cm}|p{0.8cm}|p{0.8cm}|p{0.8cm}|p{0.8cm}|}
\hline
Loop&0 & 1 & 2 & 3 & 4\\\hline
Number &2 & 3 & 5 & 8 & 11 \\
\hline
\end{tabular}
\end{table}

 In Section 2, we list some basic results about $\son$ group and present the trace basis and $L$-loop color decomposition of four-point amplitudes. In Section 3, loop-level group-theoretic constraint relations among four-point color-ordered amplitudes in $\son$ theories are derived up to four-loop order. In section 4, we list some basic results about $\sp2n$ group and present the trace basis and $L$-loop color decomposition of four-point amplitudes. In section 5, loop-level constraint relations are derived for four-point amplitudes in $\sp2n$ gauge theories up to four-loop order. The last section is devoted to conclusion and discussion. Some calculation details are provided in the Appendix.

\section{Trace basis of four-point amplitudes in $\son$ gauge theories}
We first list some basic results and our conventions about $\son$ group and algebra. In the fundamental representation, the generators of $\son$ algebra are antisymmetric, traceless matrices, which can be chosen as follows,
 \ba
 &&l_{ij}=-i(e_{ij}-e_{ji}),i,j=1,2,...,N,\nl
 &&(e_{ij})_{kl}=\d_{ik}\d_{jl}.
 \ea
The above generators can be denoted by $\{T^a\}$ ($a=1,2,...,N(N-1)/2$). They satisfy the $\son$ Lie algebra
 \ba\label{algebra}
 \[T^a, T^b\]=if^{abc}T^c,
 \ea
where $f^{abc}$ is the $\son$ structure constant, and the normalization condition
 \ba\label{normalization}
 \tr(T^a T^b)=2\d^{ab}.
 \ea
 From the above two equations, the structure constant can be expressed as trace of generators
 \ba\label{ftot}
 f^{abc}=-\frac{i}{2}\tr([T^a,T^b]T^c)=-i\tr(T^a T^b T^c).
 \ea
The quadratic Casimir of $\son$ is
 \ba\label{casimir}
 \sum_a T^a T^a=(N-1)I_{N\times N}.
 \ea
Two useful identities for the trace of $\son$ generators are
 \ba\label{iden}
 &&\tr(T^a A T^a B)=\tr(A)\tr(B)-(-1)^{n_B}(AB^r),\\
 &&\tr(T^a A)\tr(T^a B)=\tr(AB)-(-1)^{n_B}(AB^r),
 \ea
where $A,B$ are products of $\son$ generator matrices or $I_{N\times N}$. $n_{B}$ is the number of generators in $B$ and $n_{B}=0$ if $B=I_{N\times N}$.
If $B=T^1 T^2 T^3\cdots T^{n-1}T^{n}$, then $B^r=T^{n}T^{n-1}\cdots T^3 T^2 T^1$.

It is known that scattering amplitudes in $\sun$ gauge theories can be decomposed into two parts: color parts and kinematic parts. The color parts can be expressed in terms of structure constants or traces of $\sun$ matrices in the fundamental representation. This leads to two kinds of decomposition, which can be called $f$-based decomposition and trace-based decomposition. In $f$-based decomposition \cite{ddm2000,DelDuca:1999iql}, a full $n$-point amplitude is written as
 \ba\label{fdecom}
 \cA_n=\sum_\l c_\l a_\l,
 \ea
 where a color factor $c_\l$ is a product of several structure constructs and $a_\l$ is the corresponding kinematic factor. In trace-based decomposition, the full $n$-particle amplitude is written as
 \ba\label{tdecom}
 \cA_n=\sum_i t_i A_i,
 \ea
 where the trace basis $\{t_i\}$ includes independent traces of $\sun$ matrices and $A_i$ is the color-ordered amplitude corresponding to $t_i$ . Using some identities of $\sun$ algebra, a color factor $c_\r$ can be written as a linear combination of trace basis elements,
 \ba
 c_\r=\sum_i M_{\r i}t_i.
 \ea
 Then, the color-ordered amplitude $A_i$ can be expressed as
 \ba
 A_i=\sum_\r a_\r M_{\r i}.
 \ea
If the matrix $M$ has a nonzero right null vector $r$, then we can obtain a constraint equation for the color-ordered amplitudes, $A\cdot r=0$.

The above is the basic idea used before to derive group-theoretic constraints on color-ordered amplitudes.  We adopt the same idea to derive constraints on four-point color-ordered amplitudes in $\son$ gauge theories. The trace basis for four-point amplitudes in $\son$ gauge theory is
 \ba\nonumber
 T_1=\tr(T^{a_1}T^{a_2}T^{a_3}T^{a_4}) ~~~&~~~  T_4=\tr(T^{a_1}T^{a_2})\tr(T^{a_3}T^{a_4}) \\\nonumber
 T_2=\tr(T^{a_1}T^{a_2}T^{a_4}T^{a_3}) ~~~&~~~  T_5=\tr(T^{a_1}T^{a_3})\tr(T^{a_2}T^{a_4}) \\
 T_3=\tr(T^{a_1}T^{a_4}T^{a_2}T^{a_3}) ~~~&~~~  T_6=\tr(T^{a_1}T^{a_4})\tr(T^{a_2}T^{a_3})
 \ea
At tree level, only single traces appear in the trace-based color decomposition. At one-loop order, double-trace terms appear. At higher-loop order, traces with different powers of $N$ appear in the decomposition. Thus, the trace-based color decomposition of four-point $L$-loop amplitude has the following form
 \ba\label{decomp}
 \cA^L=\sum_{m=0}^L\sum_{i=1}^3 N^{m}T_i A_i^{(L,m)} +\sum_{n=0}^{L-1}\sum_{j=4}^6 N^{n}T_j A_j^{(L,n)}.
 \ea
$\{A_i^{(L,m)}\}$ and $\{A_j^{(L,n)}\}$ are the color-ordered amplitudes at $L$-loop order. This decomposition allows for a relative factor of $N$ between the leading and subleading contributions to a given $T_{i(j)}$. This is different from the $\sun$ case where the suppression always comes in $N^2$ \cite{nacu2012}.
The reason is that the algebraic identities satisfied by $\sun$ and $\son$ have different structural features.
For $\son$, both single-trace term and double-trace term appear on the right side of the identity \eqref{iden}.
For $\sun$, similar identity is $\tr(T^a A T^a B)=\tr(A)\tr(B)-\frac{1}{N}\Tr(AB)$. When we only consider adjoint particles, $\frac{1}{N}$ terms always cancel and can be omitted. Thus, only double-trace term appears on the right side of the $\sun$ identity. Using these identities to construct higher-loop trace basis
from tree-level single-trace basis, we can obtain the different suppression behaviors between $\son$ and $\sun$.

\section{constraint relations among $\son$  four-point amplitudes}
As discussed in the previous section, in order to find the constraint relations among four-point $L$-loop color-ordered amplitudes, we should first find out a complete color basis for the $L$-loop amplitudes, and then find the transformation matrix $M$ between the color basis and $L$-loop trace basis. The right null vectors of $M$ give the constraint relations.

At tree level, there are three kinds of color factors
 \ba
 c_s=f^{a_1 a_2 b}f^{b a_3 a_4},~~~~~~c_t=f^{a_4 a_1 b}f^{b a_2 a_3},~~~~~~c_u=f^{a_3 a_1 b}f^{b a_2 a_4},
 \ea
which correspond to $s,t,u$-channel diagrams respectively. These color factors are not independent and satisfy Jocobi identity: $c_s=c_t-c_u$. So we can choose $c_s,c_t$ as independent color basis elements. According to eqs.\eqref{ftot} and \eqref{iden}, $c_s,c_t$ can be expressed by tree-level trace basis $\{T_i,i=1,2,3\}$. The corresponding transformation matrix is
 \ba\label{zero}
 M^{(0)}=\left(
       \begin{array}{ccc}
         -1 & 1 & 0 \\
         -1 & 0 & 1 \\
       \end{array}
     \right).
 \ea
The right null vector of $M^{(0)}$ is
 \ba\label{r0}
 r^{(0)}=(1,1,1)',
 \ea
where the prime means transpose of matrix. This vector implies a relation among the color-ordered amplitudes,
 \ba
 A^{(0,0)}(1,2,3,4)+A^{(0,0)}(1,2,4,3)+A^{(0,0)}(1,4,2,3)=0.
 \ea
This relation for the $\son$ amplitudes is the same as that for $\sun$ amplitudes.

For $L$-loop amplitudes, it is complicated to explicitly find out a complete color basis. Here we adopt the same assumption used in \cite{nacu2012} that all $(L+1)$-loop color factors can be obtained from $L$-loop color factors by attaching a rung between two of its external legs. This assumption can be checked explicitly at lower loop orders ($L\leqslant 4$) for $SU(N)$ \cite{nacu2012} and is thought to be correct at $L>4$ loop orders. For a diagram with color factor
 \ba
c=f^{* a_i *}\cdots f^{*a_j*}\cdots,
 \ea
where "$\cdots$" denotes product of some structure constants. Attaching a rung between its external legs $(i,j)$ means the color factor of the resulting diagram is
 \ba\label{rung}
\tilde{c}=f^{b a_i e}f^{e a_j d}f^{* b *}\cdots f^{*d*}\cdots.
 \ea

For four-point amplitudes, we assume that all possible color factors of $(L+1)$-loop diagrams can be obtained from $L$-loop color factors by attaching legs $(1,2),(1,3),(1,4)$ as in eq.\eqref{rung}. Then we consider the effect of this attaching process on the trace basis. Let $ T=\left(
                                                                                                  \begin{array}{c}
                                                                                                    T_1 \\
                                                                                                    T_2\\
                                                                                                    T_3\\
                                                                                                  \end{array}
                                                                                                \right)
$, $\tilde{T}=\left(
                \begin{array}{c}
                  T_4 \\
                  T_5\\
                  T_6\\
                \end{array}
              \right)
$. After the process of attaching a rung, $T$ and $\tilde{T}$ transform as
 \ba\label{TraceTrans}
 T\rightarrow \(A,B,C\)\left(
                         \begin{array}{c}
                           N T \\
                           T\\
                           \tilde{T} \\
                         \end{array}
                       \right),~~~~~~
 \tilde{T}\rightarrow\(D,E,F\)\left(
                                \begin{array}{c}
                                  N \tilde{T} \\
                                  T\\
                                  \tilde{T} \\
                                \end{array}
                              \right),
 \ea
where $A, B, C, D, E, F$ are all $3\times 3$ matrices and are different for each attaching way. These matrices are given explicitly in the Appendix.

Let us explain how to use the iterative method to derive higher-loop right null vectors from lower-loop ones. Suppose $\{c^{(L)}_\a\}$ is a complete (maybe overcomplete) color basis for $L$-loop amplitudes, and they can be expanded by $L$-loop trace basis
$\{T^{(L)}_k,(k=1,2,\cdots,6L+3)\}=\{N^L T_i,N^{L-1} T_i,N^{L-1} T_j,\cdots, T_i, T_j, (i=1,2,3,j=4,5,6)\}$,
 \ba\label{Mjuzhen}
 c_\a^{(L)}=\sum_{k=1}^{6L+3} M_{\a k}^{(L)}T_k^{(L)}.
 \ea
 A $L$-loop four-point full amplitude can be written as
 \ba\label{decomp1}
 \cA^L=\sum_{k=1}^{6L+3}T_k^{(L)}A_k^{(L)}=\sum_\a c_\a^{(L)}a_\a^{(L)},
 \ea
 where $\{A_k^{(L)}\}$ is one-to-one corresponding to $\{A^{(L,m)}\}$ in \eqref{decomp}.
A $L$-loop right null vector $r^{(L)}$ of $M^{(L)}$ satisfies
 \ba
 \sum_{k=1}^{6L+3} M_{\a k}^{(L)}r^{(L)}_k=0,
 \ea
and implies a constraint on $L$-loop color-ordered amplitudes
 \ba
\sum_{k=1}^{6L+3} A^{(L)}_k r^{(L)}_k=0.
 \ea
The $(L+1)$-loop color factors can be obtained by the attaching procedure and this procedure also transforms a $L$-loop trace basis element to a linear combination of $(L+1)$-loop trace basis,
 \ba
 T_k^{(L)}\rightarrow \sum_{l=1}^{6L+9} G_{kl}^{(L,L+1)}T_l^{(L+1)},
 \ea
where $G$ is a $(6L+3)\times(6L+9)$ transformation matrix.
Then we have
 \ba
 c_\a^{(L+1)}=\sum_{l=1}^{6L+9} M_{\a l}^{(L+1)}T_l^{(L+1)}=\sum_{k=1}^{6L+3}\sum_{l=1}^{6L+9} M_{\a k}^{(L)}G_{kl}^{(L,L+1)}T_l^{(L+1)}.
 \ea
A $(L+1)$-loop right null vector $ r^{(L+1)}$ must satisfy
 \ba
\sum_{l=1}^{6L+9} M_{\a l}^{(L+1)}r^{(L+1)}_l=\sum_{k=1}^{6L+3}\sum_{l=1}^{6L+9} M_{\a k}^{(L)}G_{kl}^{(L,L+1)}r^{(L+1)}_l=0,
 \ea
which means
 \ba\label{nuvere}
 G^{(L,L+1)}\cdot r^{(L+1)}=\textrm{linear combination of } \{r^{(L)}\}.
 \ea
  This is the relation between $L$-loop and $(L+1)$-loop right null vectors and is the basic equation of the recursive method.
\subsection{constraints on one-loop amplitudes}
We have obtained $r^{(0)}$ in eq.\eqref{r0}.  The transformation matrix $G^{(0,1)}$ between trace bases of tree amplitudes and one-loop amplitudes is
 \ba
 G^{(0,1)}=(A,B,C).
 \ea
The explicit forms of $A,B,C$ are in the Appendix. By solving the basic recursive equation
 \ba\label{recursiveeq}
 G^{(0,1)}\cdot r^{(1)}=\textrm{linear combination of }\{r^{(0)}\},
 \ea
we get six one-loop right null vectors, $\{r^{(1)}\}=$
 \ba\nonumber
 &(4,2,2,1,0,0,0,0,0)',(2,4,2,0,1,0,0,0,0)',\\\nonumber
 &(2,2,4,0,0,1,0,0,0)',(-1,-1,-1,0,0,0,1,0,0)',\\
 &(-1,-1,-1,0,0,0,0,1,0)',(-1,-1,-1,0,0,0,0,0,1)'.
 \ea
The last three null vectors imply the following equations,
 \ba
 A^{(1,0)}(1,2;3,4)=A^{(1,0)}(1,3;2,4)=A^{(1,0)}(1,4;2,3),\\\label{1L2}
  A^{(1,0)}(1,2;3,4)= A^{(1,1)}(1,2,3,4)+ A^{(1,1)}(1,2,4,3)+ A^{(1,1)}(1,4,2,3).
 \ea
 The three color-ordered amplitudes corresponding to double-trace basis elements are equal to each other and they are linear combinations of leading-order amplitudes. The other three vectors imply the following constraint equations
 \ba
 4A^{(1,1)}(1,2,3,4)+2A^{(1,1)}(1,2,4,3)+2A^{(1,1)}(1,4,2,3)=-A^{(1,0)}(1,2,3,4),\\
 2A^{(1,1)}(1,2,3,4)+4A^{(1,1)}(1,2,4,3)+2A^{(1,1)}(1,4,2,3)=-A^{(1,0)}(1,2,4,3),\\\label{1L6}
 2A^{(1,1)}(1,2,3,4)+2A^{(1,1)}(1,2,4,3)+4A^{(1,1)}(1,4,2,3)=-A^{(1,0)}(1,4,2,3).
 \ea
 From \eqref{1L2} to \eqref{1L6}, all sub-leading color-ordered amplitudes are expressed as linear combinations of three leading-order amplitudes.
So, at one-loop order, the number of independent amplitudes is 3..

 The number of independent right null vectors can also be obtained from direct counting of independent color factors and trace basis elements at one-loop level. It is known that there are 3 independent one-loop color  factors corresponding to 3 different box diagrams and there are 9 ($6L+3$) independent trace basis elements. Thus the number of independent null vectors should be 6. This is consistent with the recursive result.
\subsection{constraints on two-loop amplitudes}
In this section we derive the constraints on two-loop amplitudes. The transformation matrix between one-loop and two-loop trace bases is
\ba
G^{(1,2)}= \left(
   \begin{array}{ccccc}
     A & B & C & 0 & 0 \\
     0 & A & 0 & B & C \\
     0 & 0 & D & E & F \\
   \end{array}
 \right).
 \ea
By solving recursive equation
 \ba
 G^{(1,2)}\cdot r^{(2)}=\textrm{linear combination of }\{r^{(1)}\},
 \ea
we get 10 two-loop right null vectors, $\{r^{(2)}\}=$
 \ba\nonumber
&&(10,10,10,1,1,1,0,0,0,0,0,0,0,0,0)',\\\nonumber
&&(7,7,3,1,1,0,1,0,0,0,0,0,0,0,0)',(-7,-3,-3,-1,0,0,0,1,0,0,0,0,0,0,0)',\\\nonumber
&&(-3,-7,-3,0,-1,0,0,0,1,0,0,0,0,0,0)',\\\nonumber
&&(40,12,12,8,0,0,0,0,0,1,0,0,0,0,0)',(12,40,12,0,8,0,0,0,0,0,1,0,0,0,0)',\\\nonumber
&&(-68,-68,-40,-8,-8,0,0,0,0,0,0,1,0,0,0)',\\\nonumber
&&(-30,-30,-14,-4,-4,0,0,0,0,0,0,0,1,0,0)',(26,10,10,4,0,0,0,0,0,0,0,0,0,1,0)',\\
&&(10,26,10,0,4,0,0,0,0,0,0,0,0,0,1)',
 \ea
which imply 10 constraint relations between 15 two-loop color-ordered amplitudes. So the number of independent amplitudes at two-loop order is 5.

The first null vector implies the following linear constraint on the 6 leading amplitudes
 \ba\nn
 10 (A^{(2,2)}(1,2,3,4)+A^{(2,2)}(1,2,4,3)+A^{(2,2)}(1,4,2,3))\\
 +A^{(2,1)}(1,2,3,4)+A^{(2,1)}(1,2,4,3)+A^{(2,1)}(1,4,2,3)=0.
 \ea
Then only 5 of the 6 leading amplitudes are independent. From the other null vectors, all other sub-leading color-ordered amplitudes can be expressed as linear combinations of the 6 leading amplitudes. The second null vector implies
 \ba\nn
 7 A^{(2,2)}(1,2,3,4)+7 A^{(2,2)}(1,2,4,3)+3A^{(2,2)}(1,4,2,3)\\+A^{(2,1)}(1,2,3,4)+A^{(2,1)}(1,2,4,3)+A^{(2,1)}(1,2;3,4)=0.
 \ea
The third and forth null vectors imply relations which  are equivalent to permutations of the above.
The relation implied by the seventh null vector is
 \ba\nn
 -68 A^{(2,2)}(1,2,3,4)-68 A^{(2,2)}(1,2,4,3)-40A^{(2,2)}(1,4,2,3)\\-8A^{(2,1)}(1,2,3,4)-8A^{(2,1)}(1,2,4,3)+ A^{(2,0)}(1,4,2,3)=0.
 \ea
 The fifth and sixth null vectors imply relations which are equivalent to perturbations of the above.
 The eighth null vectors give the following relation
  \ba\nn
 -30 A^{(2,2)}(1,2,3,4)-30 A^{(2,2)}(1,2,4,3)-14A^{(2,2)}(1,4,2,3)\\-4A^{(2,1)}(1,2,3,4)-4A^{(2,1)}(1,2,4,3)+A^{(2,0)}(1,2;3,4)=0.
  \ea
 The ninth and tenth null vectors imply relations which are equivalent to permutations of the above.

At two-loop order, there are 6 planar ladder diagrams. The color factors of them are not all linear independent and satisfy one linear constraint equation.
So there are 5 linearly independent planar color factors. All color factors of nonplanar diagrams can be written as linear combinations of the planar color factors. One typical color factor of a planar ladder diagram is $c^{(2,Lad)}_{1234}=f^{a_1 1 a_2}f^{a_2 2 a_3}f^{a_3 a_4 a_7}f^{a_4 3 a_5}f^{a_5 4 a_6}f^{a_6 a_1 a_7}$. One typical
color factor of a nonplanar diagram is $c^{(2,NP)}_{1234}=f^{a_1 1 a_2}f^{a_2 2 a_3}f^{a_3 a_6 a_4}f^{a_4 4 a_5}f^{a_5 a_1 a_7}f^{a_6 3 a_7}$. Two relations among planar and nonplanar color factors are
 \ba\label{so2looplad}
 &c^{(2,Lad)}_{1234}-c^{(2,Lad)}_{1432}+c^{(2,Lad)}_{1423}-c^{(2,Lad)}_{1324}+c^{(2,Lad)}_{1342}-c^{(2,Lad)}_{1243}=0,\\
 &3c^{(2,NP)}_{1234}=c^{(2,Lad)}_{1234}-c^{(2,Lad)}_{1432}-c^{(2,Lad)}_{1342}+c^{(2,Lad)}_{1243}.
 \ea
The dimension of two-loop trace basis  is 15. Thus by counting the numbers of color basis elements and trace basis elements,  10 independent right null vectors should be obtained. This is consistent with the recursive results.
\subsection{constraints on three-loop amplitudes}
The three-loop right null vectors obtained by recursive method are
 \ba\nonumber
 &&(21,21,21,2,2,2,1,1,1,0,0,0,0,0,0,0,0,0,0,0,0)',\\\nonumber
 &&(72,0,0,12,-2,-2,0,-4,0,1,0,0,0,0,0,0,0,0,0,0,0)',\\\nonumber
 &&(84,156,84,6,20,6,4,4,0,0,1,0,0,0,0,0,0,0,0,0,0)',\\\nonumber
 &&(0,0,72,-2,-2,12,-4,0,0,0,0,1,0,0,0,0,0,0,0,0,0)',\\\nonumber
 &&(-14,-14,-14,-1,-1,-1,6,0,0,0,0,0,1,0,0,0,0,0,0,0,0)',\\\nonumber
 &&(-14,-14,-14,-1,-1,-1,0,6,0,0,0,0,0,1,0,0,0,0,0,0,0)',\\\nonumber
 &&(-140,-140,-140,-13,-13,-13,-6,-6,0,0,0,0,0,0,1,0,0,0,0,0,0)',\\\nonumber
 &&(-256,-72,-72,-32,0,0,0,16,0,0,0,0,0,0,0,1,0,0,0,0,0)',\\\nonumber
 &&(-408,-592,-408,-32,-64,-32,-16,-16,0,0,0,0,0,0,0,0,1,0,0,0,0)',\\\nonumber
 &&(-72,-72,-256,0,0,-32,16,0,0,0,0,0,0,0,0,0,0,1,0,0,0)',\\\nonumber
 &&(100,100,100,8,8,8,-8,0,0,0,0,0,0,0,0,0,0,0,1,0,0)',\\\nonumber
 &&(100,100,100,8,8,8,0,-8,0,0,0,0,0,0,0,0,0,0,0,1,0)',\\
 &&(268,268,268,24,24,24,8,8,0,0,0,0,0,0,0,0,0,0,0,0,1)'.
 \ea
 These right null vectors imply 13 constraint relations to the 21 three-loop color-ordered amplitudes. So the number of independent amplitudes are 8.

 The explicit constraint equations can be written out easily from the null vectors and we don't do it here. The first vector implies a constraint on the first 9 color-ordered amplitudes $\{A_k^{(3)},k=1,2,\cdots,9\}$ in \eqref{decomp1}. The next 12 vectors mean that all other amplitudes $\{A_k^{(3)},k=10,11,\cdots,21\}$ are linear combinations of the first 9 amplitudes. These 12 vectors can be divided into 4 sets sequentially.  Three  vectors (or equivalent vectors) in one set satisfy certain permutation symmetries.  At three-loop order, the number of independent color factors is 8 \cite{Bern:2010tq}. By counting the numbers of color basis elements and trace basis elements, the number of right null vectors should be 13.
\subsection{constraints on four-loop amplitudes}
At four-loop order, the number of linearly independent color basis elements is 11 \cite{Bern:2010tq}. By counting the numbers of color basis elements and trace basis elements, the number of right null vectors should be 16. Using recursive method, we find that the number of right null vectors is exactly 16. A set of independent vectors can be chosen as
\ba\nonumber
&(242,242,242,22,22,22,2,2,2,1,1,1,0,0,0,0,0,0,0,0,0,0,0,0,0,0,0)'\\\nonumber
&(\frac{116}{5},\frac{116}{5},\frac{12}{5},3,3,\frac{-1}{5},\frac{44}{5},0,0,\frac{1}{5},\frac{1}{5},0,1,0,0,0,0,0,0,0,0,0,0,0,0,0,0)'\\\nonumber
&(-46,\frac{-126}{5},\frac{-126}{5},\frac{-23}{5},\frac{-7}{5},\frac{-7}{5},\frac{-2}{5},\frac{42}{5},\frac{-2}{5},\frac{-1}{5},
0,0,0,1,0,0,0,0,0,0,0,0,0,0,0,0,0)'\\\nonumber
&(\frac{-126}{5}       ,   -46     ,      \frac{-126}{5}      ,     \frac{-7}{5}       ,   \frac{-23}{5}    ,       \frac{-7}{5}      ,     \frac{-2}{5} , \frac{-2}{5}         ,  \frac{42}{5}         ,   0          ,   \frac{-1}{5}        ,    0          ,    0            ,  0
  , 1         ,     0      ,        0      ,        0       ,       0            ,  0              ,0,
 0       ,       0       ,       0     ,         0         ,     0              ,0 )'\\\nonumber
&(288         ,   -96        ,    -96      ,       56       ,     -12           , -12            ,  0,
 -8         ,     0          ,    8      ,        0        ,      0          ,    0             , 0
, 0         ,     1        ,      0      ,        0        ,      0             , 0              ,0
,0 ,0 ,0 ,0 ,0, 0)'\\\nonumber
&(-96          ,  288        ,    -96         ,   -12           ,  56           , -12           ,   0
   ,0            , -8             , 0             , 8             , 0              ,0              ,0
 ,0 ,0 ,1 ,0 ,0 ,0 ,0 ,0 ,0 ,0 ,0 ,0 ,0 )'\\\nonumber
&(-2032         , -2032         , -1648           ,-188           ,-188           ,-120            ,-24
 ,-16            ,-16            , -8            , -8              ,0              ,0              ,0
  ,0 ,0 ,0 ,1 ,0 ,0 ,0 ,0 ,0 ,0 ,0 ,0 ,0 )'\\\nonumber
&(\frac{-1362}{5}       , \frac{-1362}{5}        , \frac{-114}{5}          ,-34           , -34             ,\frac{22}{5}         ,\frac{-138}{5}
   ,2             , 2           , \frac{-12}{5}          ,\frac{-12}{5}            ,0              ,0              ,0
  ,0 ,0 ,0 ,0 ,1 ,0 ,0 ,0 ,0 ,0 ,0 ,0 ,0 )'\\\nonumber
&( 558           ,\frac{1542}{5}         ,\frac{1542}{5}          ,\frac{286}{5}           ,\frac{94}{5}           ,\frac{94}{5}           ,\frac{34}{5}
 ,\frac{-114}{5}           ,\frac{34}{5}           ,\frac{12}{5 }          , 0              ,0              ,0              ,0
  ,0 ,0 ,0 ,0 ,0 ,1 ,0 ,0 ,0 ,0 ,0 ,0 ,0)'\\\nonumber
&( \frac{1542}{5}          ,558 , \frac{1542}{5}, \frac{94}{5},\frac{ 286}{5},\frac{94}{5},\frac{34}{5}, \frac{34}{5}, \frac{-114}{5}, 0            , \frac{12}{5} , 0              ,0,0 ,0 ,0 ,0 ,0 ,0 ,0 ,1 ,0 ,0 ,0 ,0 ,0 ,0)'\\\nonumber
&( -1440           , \frac{304}{5},\frac{304}{5},\frac{-1088}{5}, \frac{128}{5},\frac{128}{5},\frac{-32}{5},\frac{192}{5}          ,\frac{-32}{5} ,\frac{-96}{5} , 0              ,0              ,0              ,0  ,0 ,0 ,0 ,0 ,0 ,0 ,0 ,1 ,0 ,0 ,0, 0 ,0)'\\\nonumber
&( \frac{304}{5}   , -1440           , \frac{304}{5}, \frac{128}{5} ,\frac{-1088}{5} , \frac{128}{5},\frac{-32}{5} ,\frac{-32}{5}, \frac{192}{5}           , 0            ,\frac{-96}{5}           , 0              ,0             , 0
  ,0 ,0 ,0 ,0 ,0 ,0 ,0 ,0 ,1 ,0 ,0 ,0 ,0)'\\\nonumber
&( \frac{23536}{5}        ,\frac{23536}{5}       , \frac{16032}{5} , 448           , 448           ,\frac{1024}{5} ,\frac{384}{5}, 32             ,32            , \frac{96}{5}           ,\frac{96}{5}            ,0              ,0              ,0
  ,0 ,0 ,0 ,0 ,0 ,0 ,0 ,0 ,0 ,1 ,0 ,0 ,0)'\\\nonumber
&( \frac{4112}{5}         ,\frac{4112}{5}          ,\frac{784}{5}           ,96             ,96            ,\frac{-32}{5}          ,\frac{168}{5},-8             ,-8            , \frac{32}{5}  ,\frac{32}{5}            ,0              ,0              ,0
  ,0 ,0, 0, 0 ,0 ,0 ,0 ,0 ,0 ,0 ,1 ,0 ,0)'\\\nonumber
&(-1392,\frac{-3632}{5},\frac{-3632}{5},\frac{-736}{5},\frac{-224}{5},\frac{-224}{5},\frac{-104}{5}
,\frac{104}{5},\frac{-104}{5},\frac{-32}{5} ,0,0,0,0,0 ,0 ,0 ,0 ,0 ,0 ,0 ,0 ,0 ,0 ,0 ,1 ,0)'\\
&( \frac{-3632}{5},-1392,          \frac{-3632}{5},         \frac{-224}{5},\frac{-736}{5},         \frac{-224}{5},\frac{-104}{5} ,\frac{-104}{5},\frac{104}{5},0,\frac{-32}{5},0,0 ,0 ,0 ,0 ,0 ,0 ,0 ,0 ,0, 0, 0 ,0 ,0, 0, 1)'
\ea
These null vectors imply 16 constraint relations for 27 four-loop color-ordered amplitudes. So there are 11 independent amplitudes. The first vector implies a constraint on the first 12 color-ordered amplitudes $\{A_k^{(4)},k=1,2,\cdots,12\}$ in \eqref{decomp1}. The next 15 vectors mean that all other amplitudes $\{A_k^{(4)},k=13,14,\cdots,27\}$ are linear combinations of the first 12 amplitudes. Again, these 15 relations can be divided into five sets. In each set, three relations (or equivalent relations) satisfy certain permutation symmetries.

\section{Trace basis of four-point amplitudes in $\sp2n$ gauge theories}
In this section, basic results and conventions about $\sp2n$ group and algebra are presented. In the fundamental representation, the generators of $\sp2n$ algebra are antisymmetric, Hermitian matrices. $2N$ row and column indices are denoted by $1,\bar{1},2,\bar{2},...,N,\bar{N}$.  The $N(2N+1)$ generators can be chosen as follows,
 \ba\nonumber
 &&-\frac{i}{\sqrt{2}}(e_{ij}+e_{\bar{i}\bar{j}}-e_{ji}-e_{\bar{j}\bar{i}}),
 \frac{1}{\sqrt{2}}(e_{i\bar{j}}+e_{\bar{i}j}+e_{j\bar{i}}+e_{\bar{j}i}),\\\nn
 &&-\frac{i}{\sqrt{2}}(e_{i\bar{j}}-e_{\bar{i}j}+e_{j\bar{i}}-e_{\bar{j}i}),
 \frac{1}{\sqrt{2}}(e_{ij}-e_{\bar{i}\bar{j}}+e_{ji}-e_{\bar{j}\bar{i}}),\\
 &&(e_{k\bar{k}}+e_{\bar{k}k}), ~~~i(e_{\bar{k}k}-e_{k\bar{k}}), ~~~(e_{kk}-e_{\bar{k}\bar{k}}),
 \ea
 where $1\leqslant i<j\leqslant N, 1\leqslant k \leqslant N$. Entries of the $e$-matrix are $(e_{\a\b})_{\r\s}=\d_{\a\r}\d_{\b\s}$$(\a,\b,\r,\s=1,\bar{1},...,N, \bar{N})$. These generators can be denoted by $\{T^a\}$ ($a=1,2,...,N(2N+1)$). Here and after, for simplicity, we use the same indices ${a,b,...}$ as the $\son$ case. These generators satisfy the  $\sp2n$ Lie algebra
 \ba\label{sp2nalgebra}
 \[T^a, T^b\]=iF^{abc}T^c,
 \ea
where $F^{abc}$ is the $\sp2n$ structure constant. The normalization condition is
 \ba\label{sp2nnormalization}
 \tr(T^a T^b)=2\d^{ab}.
 \ea
 The structure constant can be expressed as trace of generators
 \ba\label{sp2nFtoTr}
 F^{abc}=-\frac{i}{2}\tr(T^a[T^b,T^c]).
 \ea
The quadratic Casimir is
 \ba\label{sp2ncasimir}
 \sum_a T^a T^a=(2N+1)I_{2N\times 2N}.
 \ea
Two useful identities for the trace of the $\sp2n$ generators in the fundamental representation are
 \ba\label{sp2niden}\nonumber
 &&\tr(T^a A T^a B)=\tr(A)\tr(B)+(-1)^{n_B}(AB^r),\\
 &&\tr(T^a A)\tr(T^a B)=\tr(AB)-(-1)^{n_B}(AB^r),
 \ea
where $A,B$ are products of series of $\sp2n$ generator matrices or $I_{2N\times 2N}$. $n_{B}$ is the number of generators in $B$ and $n_{B}=0$ if $B=I_{2N\times 2N}$.

The trace basis for four-point tree amplitudes in $\sp2n$ gauge theories are
 \ba\nonumber
 T_1=\tr(T^{a_1}T^{a_2}T^{a_3}T^{a_4}), ~~~&~~~  T_4=\tr(T^{a_1}T^{a_2})\tr(T^{a_3}T^{a_4}), \\\nonumber
 T_2=\tr(T^{a_1}T^{a_2}T^{a_4}T^{a_3}), ~~~&~~~  T_5=\tr(T^{a_1}T^{a_3})\tr(T^{a_2}T^{a_4}), \\
 T_3=\tr(T^{a_1}T^{a_4}T^{a_2}T^{a_3}), ~~~&~~~  T_6=\tr(T^{a_1}T^{a_4})\tr(T^{a_2}T^{a_3}).
 \ea
 The full four-point $L$-loop $\sp2n$ gauge amplitude $\bar{\cA}^L$ can be decomposed in $L$-loop trace basis,
 \ba\label{sp2ndecomp}
 \bar{\cA}^L=\sum_{m=0}^L\sum_{i=1}^3 N^{m}T_i \bA_i^{(L,m)} +\sum_{n=1}^{L-1}\sum_{j=4}^6 N^{n}T_j \bA_j^{(L,n)}.
 \ea
$\{\bA_i^{(L,m)}\}$ and $\{\bA_j^{(L,n)}\}$ are the color-ordered $L$-loop amplitudes in $\sp2n$ case.
The above decomposition can also  be written in another useful form,
 \ba\label{sp2ndecomp1}
\bar{\cA}^L=\sum_{k=1}^{6L+3}T^{(L)}_k \bA^{(L)}_k,
 \ea
where $\{T^{(L)}_k,k=1,2,\cdots,6L+3\}=\{N^L T_i,N^{L-1} T_i,N^{L-1} T_j,\cdots, T_i, T_j, (i=1,2,3,j=4,5,6)\}$.

\section{constraint relations among $\sp2n$ four-point amplitudes}
In this section, we use the same procedure as $\son$ case to derive group-theoretic relations for four-point amplitudes in $\sp2n$ up to four-loop level.
At tree level, there are three kinds of color factors $c_s,c_t,c_u$. They have the  same forms as the  $\son$ case and satisfy Jocobi identity: $c_s=c_t-c_u$.
The transformation matrix between color factors and trace basis is also the same as $\son$ case. Then at tree level,
the right null vector is the same as \eqref{r0},
 \ba\label{sp2nr0}
 \bar{r}^{(0)}=(1,1,1)'.
 \ea

 The loop-level right null vectors for $\sp2n$ can be derived as $\son$ case. It is found that the transformation matrices $\{G^{(L,L+1)}\}$ of $\sp2n$ have similar structures as $\son$ case. The $A,B,C,D,E,F$ matrices in \eqref{TraceTrans} of $\sp2n$ are different from those of $\son$.
 The numbers of right null vectors  are the same for both cases at each loop order. This is consistent with the result that the numbers of independent color basis elements are the same for all simple groups at each loop order( at least up to four loop) \cite{Bern:2010tq,nacu2012}.
\subsection{constraints on one-loop amplitudes}
Using the recursive method, the independent one-loop right null vectors are
 \ba\label{sp2nr1}\nn
 &(-2,-1,-1,1,0,0,0,0,0)',~~~
 (-1,-2,-1,0,1,0,0,0,0)',\\\nn
 &(-1,-1,-2,0,0,1,0,0,0)',~~~
(\frac{-1}{2},\frac{-1}{2},\frac{-1}{2},0,0,0,1,0,0)',\\
 &(\frac{-1}{2},\frac{-1}{2},\frac{-1}{2},0,0,0,0,1,0)',~~~
(\frac{-1}{2},\frac{-1}{2},\frac{-1}{2},0,0,0,0,0,1)'.
 \ea
 The explicit constraint equations are easy to written down and are not listed here. At one-loop order, it is known that there are 3 independent color factors corresponding to 3 different box diagrams for four-point amplitudes. The number of independent trace basis elements is 9(6L+3). So, it is  necessary that 6 null vectors are obtained by recursive method. The number of independent amplitudes is 3.
\subsection{constraints on two-loop amplitudes}
From \eqref{nuvere}, the independent two-loop right null vectors for $\sp2n$ four-point color-ordered amplitudes  are
 \ba\label{sp2nr2}\nn
 (-5,-5,-5,1,1,1,0,0,0,0,0,0,0,0,0)',& (\frac{7}{2},\frac{7}{2},\frac{3}{2},-1,-1,0,1,0,0,0,0,0,0,0,0)',\\\nn
 (\frac{-7}{2},\frac{-3}{2},\frac{-3}{2},1,0,0,0,1,0,0,0,0,0,0,0)',&(\frac{-3}{2},\frac{-7}{2},\frac{-3}{2},0,1,0,0,0,1,0,0,0,0,0,0)',\\\nn
 (10,3,3,-4,0,0,0,0,0,1,0,0,0,0,0)',&(3,10,3,0,-4,0,0,0,0,0,1,0,0,0,0)',\\\nn
  (-17,-17,-10,4,4,0,0,0,0,0,0,1,0,0,0)',&(\frac{15}{2},\frac{15}{2},\frac{7}{2},-2,-2,0,0,0,0,0,0,0,1,0,0)',\\
  (\frac{-13}{2},\frac{-5}{2},\frac{-5}{2},2,0,0,0,0,0,0,0,0,0,1,0)',&(\frac{-5}{2},\frac{-13}{2},\frac{-5}{2},0,2,0,0,0,0,0,0,0,0,0,1)'.
 \ea
These vectors imply 10 constraint relations among 15 color-ordered two-loop amplitudes. So, the number of independent amplitudes is 5. At two-loop order, there are six ladder diagrams. A typical color factor of ladder diagram is
$\bar{c}^{(2,Lad)}_{1234}=F^{a1b}F^{b2c}F^{cde}F^{d3f}F^{f4g}F^{gae}$. The color factors of these ladder diagrams are not independent and satisfy a linear equation. A typical nonplanar color factor is
$\bar{c}_{1234}^{(2,NP)}=F^{a1b}F^{b2c}F^{cde}F^{d3f}F^{e4g}F^{fga}$. All the color factors of nonplanar diagrams can be written as linear combinations of those of ladder diagrams. As $\son$ case, two similar relations among planar and nonplanar color factors are
 \ba\label{sp2n2lre}
 &\bar{c}^{(2,Lad)}_{1234}-\bar{c}^{(2,Lad)}_{1432}+\bar{c}^{(2,Lad)}_{1423}-\bar{c}^{(2,Lad)}_{1324}+\bar{c}^{(2,Lad)}_{1342}-\bar{c}^{(2,Lad)}_{1243}=0,\\
 &3\bar{c}^{(2,NP)}_{1234}=\bar{c}^{(2,Lad)}_{1234}-\bar{c}^{(2,Lad)}_{1432}-\bar{c}^{(2,Lad)}_{1342}+\bar{c}^{(2,Lad)}_{1243}.
 \ea
 \subsection{constraints on three-loop amplitudes}
The 13 independent three-loop right null vectors for $\sp2n$ four-point color-ordered amplitudes are
 \ba\label{sp2nr3}\nn
 &(\frac{21}{2},\frac{21}{2},\frac{21}{2},-2,-2,-2,1,1,1,0,0,0,0,0,0,0,0,0,0,0,0)',\\\nn
 &(18,0,0,-6,1,1,0,-2,0,1,0,0,0,0,0,0,0,0,0,0,0)',\\\nn
 &(21,39,21,-3,-10,-3,2,2,0,0,1,0,0,0,0,0,0,0,0,0,0)',\\\nn
 &(0,0,18,1,1,-6,-2,0,0,0,0,1,0,0,0,0,0,0,0,0,0)',\\\nn
 &(\frac{7}{2},\frac{7}{2},\frac{7}{2},\frac{-1}{2},\frac{-1}{2},\frac{-1}{2},-3,0,0,0,0,0,1,0,0,0,0,0,0,0,0)',\\\nn
 &(\frac{7}{2},\frac{7}{2},\frac{7}{2},\frac{-1}{2},\frac{-1}{2},\frac{-1}{2},0,-3,0,0,0,0,0,1,0,0,0,0,0,0,0)',\\\nn
 &(35,35,35,\frac{-13}{2},\frac{-13}{2},\frac{-13}{2},3,3,0,0,0,0,0,0,1,0,0,0,0,0,0)',\\\nn
 &(32,9,9,-8,0,0,0,-4,0,0,0,0,0,0,0,1,0,0,0,0,0)',\\\nn
 &(51,74,51,-8,-16,-8,4,4,0,0,0,0,0,0,0,0,1,0,0,0,0)',\\\nn
 &(9,9,32,0,0,-8,-4,0,0,0,0,0,0,0,0,0,0,1,0,0,0)',\\\nn
 &(\frac{25}{2},\frac{25}{2},\frac{25}{2},-2,-2,-2,-2,0,0,0,0,0,0,0,0,0,0,0,1,0,0)',\\\nn
 &(\frac{25}{2},\frac{25}{2},\frac{25}{2},-2,-2,-2,0,-2,0,0,0,0,0,0,0,0,0,0,0,1,0)',\\
 &(\frac{67}{2},\frac{67}{2},\frac{67}{2},-6,-6,-6,2,2,0,0,0,0,0,0,0,0,0,0,0,0,1)'.
 \ea
The first vector implies a constraint on the first 9 color-ordered amplitudes $\{\bar{A}_k^{(3)},k=1,2,\cdots,9\}$ in \eqref{sp2ndecomp1}. The next 12 vectors mean that all other color-ordered amplitudes $\{\bar{A}_k^{(3)},k=10,11,\cdots,21\}$ are linear combinations of the first 9 amplitudes. So, the number of independent color-ordered amplitudes is 8. These 12 vectors can be put into 4 sets.  Three vectors (or equivalent vectors) in one set satisfy certain permutation symmetries.
\subsection{constraints on four-loop amplitudes}
 The 16 independent four-loop right null vectors obtained from \eqref{nuvere} are as follows,
 \ba\label{sp2nr4}\nn
 &(\frac{121}{2},\frac{121}{2},\frac{121}{2},-11,-11,-11,1,1,1,1,1,1,0,0,0,0,0,0,0,0,0,0,0,0,0,0,0)',\\\nn
 &(\frac{-29}{5},\frac{-29}{5},\frac{-3}{5},\frac{3}{2},\frac{3}{2},\frac{-1}{10},\frac{-22}{5},0,0,\frac{-1}{5},\frac{-1}{5},0,1,0,0,0,0,0,0,0,0,0,0,0,0,0,0)',\\\nn
 &(\frac{23}{2},\frac{63}{10},\frac{63}{10},\frac{-23}{10},\frac{-7}{10},\frac{-7}{10},\frac{1}{5},\frac{-21}{5},\frac{1}{5},\frac{1}{5},0,0,0,1,0,0,0,0,0,0,0,0,0,0,0,0,0)',\\\nn
 &(\frac{63}{10},\frac{23}{2},\frac{63}{10},\frac{-7}{10},\frac{-23}{10},\frac{-7}{10},\frac{1}{5},\frac{1}{5},\frac{-21}{5},0,\frac{1}{5},0,0,0,1,0,0,0,0,0,0,0,0,0,0,0,0)',\\\nn
 &(-36,12,12,14,-3,-3,0,2,0,-4,0,0,0,0,0,1,0,0,0,0,0,0,0,0,0,0,0)',\\\nn
 &(12,-36,12,-3,14,-3,0,0,2,0,-4,0,0,0,0,0,1,0,0,0,0,0,0,0,0,0,0)',\\\nn
 &(254,254,206,-47,-47,-30,6,4,4,4,4,0,0,0,0,0,0,1,0,0,0,0,0,0,0,0,0)',\\\nn
 &(\frac{-681}{20},\frac{-681}{20},\frac{-57}{20},\frac{17}{2},\frac{17}{2},\frac{-11}{10},\frac{-69}{10},\frac{1}{2},\frac{1}{2},\frac{-6}{5},\frac{-6}{5},0,0,0,0,0,0,0,1,0,0,0,0,0,0,0,0)',\\\nn
 &(\frac{279}{4},\frac{771}{20},\frac{771}{20},\frac{-143}{10},\frac{-47}{10},\frac{-47}{10},\frac{17}{10},\frac{-57}{10},\frac{17}{10},\frac{6}{5},0,0,0,0,0,0,0,0,0,1,0,0,0,0,0,0,0)',\\\nn
 &(\frac{771}{20},\frac{279}{4},\frac{771}{20},\frac{-47}{10},\frac{-143}{10},\frac{-47}{10},\frac{17}{10},\frac{17}{10},\frac{-57}{10},0,\frac{6}{5},0,0,0,0,0,0,0,0,0,1,0,0,0,0,0,0)',\\\nn
 &(-90,\frac{19}{5},\frac{19}{5},\frac{136}{5},\frac{-16}{5},\frac{-16}{5},\frac{-4}{5},\frac{24}{5},\frac{-4}{5},\frac{-24}{5},0,0,0,0,0,0,0,0,0,0,0,1,0,0,0,0,0)',\\\nn
 &(\frac{19}{5},-90,\frac{19}{5},\frac{-16}{5},\frac{136}{5},\frac{-16}{5},\frac{-4}{5},\frac{-4}{5},\frac{24}{5},0,\frac{-24}{5},0,0,0,0,0,0,0,0,0,0,0,1,0,0,0,0)',\\\nn
 &(\frac{1471}{5},\frac{1471}{5},\frac{1002}{5},-56,-56,\frac{-128}{5},\frac{48}{5},4,4,\frac{24}{5},\frac{24}{5},0,0,0,0,0,0,0,0,0,0,0,0,1,0,0,0)',\\\nn
 &(\frac{-257}{5},\frac{-257}{5},\frac{-49}{5},12,12,\frac{-4}{5},\frac{-21}{5},1,1,\frac{-8}{5},\frac{-8}{5},0,0,0,0,0,0,0,0,0,0,0,0,0,1,0,0)',\\\nn
 &(87,\frac{227}{5},\frac{227}{5},\frac{-92}{5},\frac{-28}{5},\frac{-28}{5},\frac{13}{5},\frac{-13}{5},\frac{13}{5},\frac{8}{5},0,0,0,0,0,0,0,0,0,0,0,0,0,0,0,1,0)',\\
 &(\frac{227}{5},87,\frac{227}{5},\frac{-28}{5},\frac{-92}{5},\frac{-28}{5},\frac{13}{5},\frac{13}{5},\frac{-13}{5},0,\frac{8}{5},0,0,0,0,0,0,0,0,0,0,0,0,0,0,0,1)'.
 \ea
 These null vectors imply 16 constraint relations for four-loop color-ordered amplitudes. The number of independent amplitudes is 11. The first vector gives a constraint to the first 12 color-ordered amplitudes $\{\bar{A}_k^{(4)},k=1,2,\cdots,12\}$ in \eqref{sp2ndecomp1}. The next 15 vectors mean that all other amplitudes $\{\bar{A}_k^{(4)},k=13,14,\cdots,27\}$ are linear combinations of the first 12 amplitudes. These 15 relations can be put into five sets. In each set, three vectors  (or equivalent vectors) satisfy certain permutation symmetries.

\section{conclusion and discussion}
In this paper, we consider the group-theoretic constraint relations for four-point color-ordered $\son$ and $\sp2n$ gauge amplitudes. These explicit relations are derived  up to four-loop order. The blocks of transformation matrices $(A,\cdots,F)$ in both theories have similar structures and the tree-level constraint relations  are the same in both theories. In both theories, there are $n=1,6,10,13,16$ linear constraint relations at $L=0,1,2,3,4$ loop orders. The result also provides a check that the numbers of independent color basis elements of four-point amplitudes for $L=0,1,2,3,4$ are $2,3 ,5,8,11$ for any simple group\cite{Bern:2010tq,nacu2012}.
In principle, higher-loop ($L\geqslant 5$) constraint relations can be derived with the recursive method order by order.
Here we stop the computation at four loop. One reason is that the dimension of null vectors $(6L+3)$ become larger and larger.
Another reason is the computational accuracy problem that entries of the higher-loop null vectors have large different
orders of magnitude and some nonzero entries will be output as zeros in numerical computation.

It is worth noting that our results and the results in \cite{nacu2012,enacu20121,enacu20122} are based on the assumption that all independent high-loop color factors  can be obtained by the attaching approach from the lower-loop ones. According to our results, it is true for $\son$ and $\sp2n$ at least up to four-loop order. Then it will be interesting to prove this assumption to all orders or to provide a general way of
counting the number of independent color factors at any loop order.

{\textbf{Acknowledgements:\\}}
This work is supported by the Natural Science Foundation of Guangdong Province (No.2016A030313444).

\appendix*
\section{}
For both $\son$ and $\sp2n$ cases, the transformation matrix $G^{(L,L+1)}$ between $L$-loop trace basis and $(L+1)$-loop trace basis  is  $(6L+3)\times(6L+9)$ and has the following form
 \ba
 \left(
   \begin{array}{cccccccc}
     A & B & C & 0 & 0 & 0 & 0 & \ldots \\
     0 & A & 0 & B & C & 0 & 0 & \ldots \\
     0 & 0 & D & E & F & 0 & 0 & \ldots \\
     0 & 0 & 0 & A & 0 & B & C & \ldots \\
     0 & 0 & 0 & 0 & D & E & F & \ldots \\
     \vdots & \vdots & \vdots & \vdots & \vdots & \vdots & \vdots & \ddots \\
   \end{array}
 \right).
 \ea
$A,B,C,D,E,F$ are all $3\times 3$ matrices.

For $\son$, these matrices are of the following forms
 \ba
 A&=&\left(
       \begin{array}{ccc}
         -e_{12}-e_{14} & 0 & 0 \\
         0 & -e_{12}-e_{13} & 0 \\
         0 & 0 & -e_{13}-e_{14} \\
       \end{array}
     \right),\\
  B&=&\left(
       \begin{array}{ccc}
         3e_{12}-2e_{13}+3e_{14} & e_{12}-e_{13} & -e_{13}+e_{14} \\
         e_{12}-e_{14} & 3e_{12}+3e_{13}-2e_{14} & e_{13}-e_{14} \\
         -e_{12}+e_{14} & -e_{12}+e_{13} & -2e_{12}+3e_{13}+3e_{14} \\
       \end{array}
     \right),\\
   C&=&\left(
       \begin{array}{ccc}
         -e_{12}+e_{13} & 0 & e_{13}-e_{14} \\
         -e_{12}+e_{14} & -e_{13}+e_{14} & 0 \\
         0 & e_{12}-e_{13} & e_{12}-e_{14} \\
       \end{array}
     \right),\\
 D&=&\left(
       \begin{array}{ccc}
         -2e_{12} & 0 & 0 \\
         0 & -2e_{13} & 0 \\
         0 & 0 & -2e_{14} \\
       \end{array}
     \right),\\
 E&=&\left(
       \begin{array}{ccc}
         4e_{13}-4e_{14} & -4e_{13}+4e_{14} & 0 \\
         0 & -4e_{12}+4e_{14} & 4e_{12}-4e_{14} \\
         -4e_{12}+4e_{13} & 0 & 4e_{12}-4e_{13} \\
       \end{array}
     \right),\\
 F&=&\left(
       \begin{array}{ccc}
         4e_{12} & 0 & 0 \\
         0 & 4e_{13} & 0 \\
         0 & 0 & 4e_{14}\\
       \end{array}
     \right).
 \ea
In the above matrices, $e_{1i}$ takes one when we attach legs $(1,i)$ and otherwise takes zero.

For $\sp2n$, these matrices are
  \ba
 A&=&\left(
       \begin{array}{ccc}
         -2 e_{12}-2e_{14} & 0 & 0 \\
         0 & -2 e_{12}-2e_{13} & 0 \\
         0 & 0 & -2e_{13}-2e_{14} \\
       \end{array}
     \right),\\
  B&=&\left(
       \begin{array}{ccc}
         -3e_{12}+2e_{13}-3e_{14} & -e_{12}+e_{13} & e_{13}-e_{14} \\
         -e_{12}+e_{14} & -3e_{12}-3e_{13}+2e_{14} & -e_{13}+e_{14} \\
         e_{12}-e_{14} & e_{12}-e_{13} & 2e_{12}-3e_{13}-3e_{14} \\
       \end{array}
     \right),\\
   C&=&\left(
       \begin{array}{ccc}
         -e_{12}+e_{13} & 0 & e_{13}-e_{14} \\
         -e_{12}+e_{14} & -e_{13}+e_{14} & 0 \\
         0 & e_{12}-e_{13} & e_{12}-e_{14} \\
       \end{array}
     \right),\\
 D&=&\left(
       \begin{array}{ccc}
         -4e_{12} & 0 & 0 \\
         0 & -4e_{13} & 0 \\
         0 & 0 & -4e_{14} \\
       \end{array}
     \right),\\
 E&=&\left(
       \begin{array}{ccc}
         4e_{13}-4e_{14} & -4e_{13}+4e_{14} & 0 \\
         0 & -4e_{12}+4e_{14} & 4e_{12}-4e_{14} \\
         -4e_{12}+4e_{13} & 0 & 4e_{12}-4e_{13} \\
       \end{array}
     \right),\\
 F&=&\left(
       \begin{array}{ccc}
         -4e_{12} & 0 & 0 \\
         0 & -4e_{13} & 0 \\
         0 & 0 & -4e_{14}\\
       \end{array}
     \right).
 \ea

\end{document}